\documentclass[12pt,english]{article}
\usepackage[T1]{fontenc}
\usepackage[latin1]{inputenc}

\makeatletter

\usepackage{babel}
\makeatother
\begin{document}

\title{Via Aristotle, Leibnitz \& Mach to a fractal $D=2$ universe }

\author{D. F. Roscoe}

\maketitle
\begin{abstract}
{ Over the past two decades or so, it has become increasingly appparent
that, out to quite large distances, galaxies are distributed in a
quasi-fractal fashion with fractal dimension $D\approx2$. Whether
or not this behaviour continues onto indefinitely large scales is
a matter of live debate and is a question that can only be settled
if, at some scale, there is an unambiguous transition to homogeneity.
This point has not yet been reached and may never be.} { This paper
has been written on the basis of the tentative hypothesis that quasi-fractal
$D\approx2$ behaviour is a persistent and fundamental feature of
galaxy distribution on all scales and addresses the question of the
origins of this putative fractality.}{ Given this tentative hypothesis
then, except for the device of putting the fractal behaviour into
the initial conditions - which is to by-pass the question, there is
no obvious explanation within the framework of conventional cosmology.
So we adopt the position that \emph{fractality} is per-se a signature
that different thinking is required.}{ We find that a beautiful
solution flows deductively from elementary observations about the
world in which we live when we take seriously a view of space and
time that can be traced, via Mach, Berkeley and Leibnitz to Aristotle.
}{In summary, we find that a globally inertial space and time can
be irreducibly associated with a fractal, $D=2$, distribution of
material if we are prepared to review our understanding of what is
meant by the notion of \emph{metric} on astrophysical scales.}
\end{abstract}

\section{Introduction}

\label{sec.evidence} A basic assumption of the \textit{Standard Model}
of modern cosmology is that, on some scale, the universe is homogeneous;
however, in early responses to suspicions that the accruing data was
more consistent with Charlier's conceptions of an hierarchical universe
\cite{key-5,key-6,key-7} than with the requirements of the \textit{Standard
Model}, de Vaucouleurs \cite{key-134} showed that, within wide limits,
the available data satisfied a mass distribution law $M\approx r^{1.3}$,
whilst Peebles \cite{key-150} found $M\approx r^{1.23}$. The situation,
from the point of view of the \textit{Standard Model}, has continued
to deteriorate with the growth of the data-base to the point that,
according to Baryshev et al \cite{key-160}

\emph{...the scale of the largest inhomogeneities (discovered to date)
is comparable with the extent of the surveys, so that the largest
known structures are limited by the boundaries of the survey in which
they are detected.}  

For example, several recent redshift surveys, such as those performed
by Huchra et al \cite{key-140}, Giovanelli and Haynes \cite{key-139},
De Lapparent et al \cite{key-135}, Broadhurst et al \cite{key-157},
Da Costa et al \cite{key-136} and Vettolani et al \cite{key-153}
etc have discovered massive structures such as sheets, filaments,
superclusters and voids, and show that large structures are common
features of the observable universe; the most significant conclusion
to be drawn from all of these surveys is that the scale of the largest
inhomogeneities observed is comparable with the spatial extent of
the surveys themselves.\\
 \\
 In recent years, several quantitative analyses of both pencil-beam
and wide-angle surveys of galaxy distributions have been performed:
three recent examples are give by Joyce, Montuori \& Labini \cite{key-141}
who analysed the CfA2-South catalogue to find fractal behaviour with
$D\,$=$\,1.9\pm0.1$; Labini \& Montuori \cite{key-143} analysed
the APM-Stromlo survey to find fractal behaviour with $D\,$=$\,2.1\pm0.1$,
whilst Labini, Montuori \& Pietronero \cite{key-144} analysed the
Perseus-Pisces survey to find fractal behaviour with $D\,$=$\,2.0\pm0.1$.
There are many other papers of this nature in the literature all supporting
the view that, out to medium depth at least, galaxy distributions
appear to be fractal with $D\,$$\approx$$\,2$.\\
 \\
 This latter view is now widely accepted (for example, see Wu, Lahav
\& Rees \cite{key-26}), and the open question has become whether
or not there is a transition to homogeneity on some sufficiently large
scale. For example, Scaramella et al \cite{key-152} analyse the ESO
Slice Project redshift survey, whilst Martinez et al \cite{key-148}
analyse the Perseus-Pisces, the APM-Stromlo and the 1.2-Jy IRAS redshift
surveys, with both groups finding evidence for a cross-over to homogeneity
at large scales. In response, the Scaramella et al analysis has been
criticized on various grounds by Joyce et al \cite{key-142}.\\
 \\
 The argument has recently reduced to a question of statistics: basically,
the proponents of the fractal view argue that the statistical tools
(eg correlation function methods) widely used to analyse galaxy distributions
by the proponents of the opposite view are deeply rooted in classical
ideas of statistics and implicitly assume that the distributions from
which samples are drawn are homogeneous in the first place. Thus,
much effort is being expended developing tools appropriate to analysing
samples drawn from more general classes of populations - a general
focus being the idea that one should not discuss fractal structures
in terms of the correlation amplitude since the only meaningful quantity
is the exponent characterizing the fractal behaviour. Recent papers
arguing this general point of view are Sylos Labini \& Gabrielli \cite{key-105}
and Gabrielli \& Sylos Labini \cite{key-106}.\\
 \\
 This work of this paper is based upon the view that this putative
fractal behaviour is fundamental and persistent on all scales, and
is a signal that our conventional understanding of space \& time needs
to be reconsidered.

\subsection{A brief history of ideas of space and time\label{sub:A-brief-history}}

The conception of space as the container of material objects is generally
considered to have originated with Democritus and, for him, it provided
the stage upon which material things play out their existence - \emph{emptiness}
exists and is that which is devoid of the attribute of \emph{extendedness}
(although, interestingly, this latter conception seems to contain
elements of the opposite view upon which we shall comment later).
For Newton \cite{key-90}, an extension of the Democritian conception
was basic to his mechanics and, for him:

\emph{... absolute space, by its own nature and irrespective of anything
external, always remains immovable and similar to itself.}  

Thus, the absolute space of Newton was, like that of Democritus, the
stage upon which material things play out their existence - it had
an \emph{objective existence} for Newton and was primary to the order
of things. In a similar way, time - \emph{universal time,} an absolute
time which is the same everywhere - was also considered to possess
an objective existence, independently of space and independently of
all the things contained within space. The fusion of these two conceptions
provided Newton with the reference system - \emph{spatial coordinates}
defined at a \emph{particular time} - by means of which, as Newton
saw it, all motions could be quantified in a way which was completely
independent of the objects concerned. It is in this latter sense that
the Newtonian conception seems to depart fundamentally from that of
Democritus - if \emph{emptiness} exists and is devoid of the attribute
of \emph{extendedness} then, in modern terms, the \emph{emptiness}
of Democritus can have no \emph{metric} associated with it. But it
is precisely Newton's belief in \emph{absolute space \& time} (with
the implied virtual clocks and rods) that makes the Newtonian conception
a direct antecedant of Minkowski spacetime - that is, of an empty
space and time within which it is possible to have an internally consistent
discussion of the notion of \emph{metric.}\\
 \\
 The contrary view is generally considered to have originated with
Aristotle \cite{key-107,key-89} for whom there was no such thing
as a \emph{void} - there was only the \emph{plenum} within which the
concept of the \emph{empty place} was meaningless and, in this, Aristotle
and Leibnitz \cite{key-85} were at one. It fell to Leibnitz, however,
to take a crucial step beyond the Aristotolian conception: in the
debate of Clarke-Leibnitz (1715$\sim$1716) \cite{key-84} in which
Clarke argued for Newton's conception, Leibnitz made three arguments
of which the second was:

\emph{Motion and position are real and detectable only in relation
to other objects ... therefore empty space, a void, and so space itself
is an unnecessary hypothesis.} 

That is, Leibnitz introduced a \emph{relational} concept into the
Aristotolian world view - what we call \emph{space} is a projection
of \emph{relationships} between material bodies into the perceived
world whilst what we call \emph{time} is the projection of ordered
\emph{change} into the perceived world. Of the three arguments, this
latter was the only one to which Clarke had a good objection - essentially
that \emph{accelerated motion,} unlike uniform motion, can be percieved
\emph{without} reference to external bodies and is therefore, he argued,
necessarily percieved with respect to the \emph{absolute space} of
Newton. It is of interest to note, however, that in rebutting this
particular argument of Leibnitz, Clarke, in the last letter of the
correspondence, put his finger directly upon one of the crucial consequences
of a relational theory which Leibnitz had apparently not realized
(but which Mach much later would) stating as absurd that:

... \emph{the parts of a circulating body (suppose the sun) would
lose the} vis centrifuga \emph{arising from their circular motion
if all the extrinsic matter around them were annihilated.} 

This letter was sent on October 29th 1716 and Leibnitz died on November
14th 1716 so that we were never to know what Leibnitz's response might
have been.\\
 \\
 Notwithstanding Leibnitz's arguments against the Newtonian conception,
nor Berkeley's contemporary criticisms \cite{key-86}, which were
very similar to those of Leibnitz and are the direct antecedants of
Mach's, the practical success of the Newtonian prescription subdued
any serious interest in the matter for the next 150 years or so until
Mach himself picked up the torch. In effect, he answered Clarke's
response to Leibnitz's second argument by suggesting that the \emph{inertia}
of bodies is somehow induced within them by the large-scale distribution
of material in the universe:

\emph{... I have remained to the present day the only one who insists
upon referring the law of inertia to the earth and, in the case of
motions of great spatial and temporal extent, to the fixed stars ...}
\cite{key-78} 

thereby generalizing Leibnitz's conception of a relational universe.
Mach was equally clear in expressing his views about the nature of
time: in effect, he viewed \emph{time} (specifically Newton's \emph{absolute
time}) as a meaningless abstraction. All that we can ever do, he argued
in \cite{key-78}, is to measure \emph{change} within one system against
\emph{change} in a second system which has been defined as the standard
(eg it takes half of one complete rotation of the earth about its
own axis to walk thirty miles).\\
 \\
 Whilst Mach was clear about the origins of inertia (in the fixed
stars), he did not hypothesize any mechanism by which this conviction
might be realized and it fell to others to make the attempt - a typical
(although incomplete) list might include the names of Einstein \cite{key-87},
Sciama \cite{key-83}, Hoyle \& Narlikar and Sachs \cite{key-81,key-82}
for approaches based on canonical ideas of spacetime, and the names
of Ghosh \cite{key-80} and Assis \cite{key-79} for approaches based
on quasi-Newtonian ideas.\\
 \\
 It is perhaps one of the great ironies of 20thC science that Einstein,
having coined the name \emph{Mach's Principle} for Mach's original
suggestion and setting out to find a theory which satisfied the newly
named Principle, should end up with a theory which, whilst albiet
enormously successful, is more an heir to the ideas of Democritus
and Newton than to the ideas of Aristotle, Leibnitz and Berkeley.
One only has to consider the special case solution of Minkowski spacetime,
which is empty but metrical, to appreciate this fact.

\section{Overview\label{sec:Overview}}

Following in the tradition of Aristotle, Leibnitz, Berkeley and Mach
we argue that no consistent cosmology should admit the possibility
of an internally consistent discussion of \emph{empty} metrical space
\& time - unlike, for example, General Relativity which has the empty
spacetime of Minkowski as a particular solution. Recognizing that
the most simple space \& time to visualize is one which is everywhere
inertial (which, in any case, approximates the reality of our universe
on very large scales) then our worldview is distilled into the elemental
question:

\emph{Is it possible to conceive a globally inertial space \& time
which is irreducibly associated with a non-trivial global mass distribution
and, if so, what are the properties of this distribution?} 

In pursuit of this question, this paper falls naturally into three
main parts, the first being a development of a Leibnitzian view of
physical space, the second being a development of what `time' means
in such a physical space and the third being a brief discussion of
dynamics in the resulting Leibnitzian universe.

\subsection{Leibnitzian physical space: overview}

We take the general position of Leibnitz about the relational nature
of space to be self-evident and in \S\ref{sec:From-Leibnitz-to},
\S\ref{sec:The-metric-tensor} and \S\ref{sec:An-invariant-calibration}
develop a quantitative Leibnitzian model of a metric three-space -
that is, of a physical three-space whose metric properties are defined
entirely in terms of its material content and for which the notion
of \emph{empty space} is a meaningless abstraction. This development
is notable for the fact that it proceeds from the simple assumption
that we can `do local physics' in the conventional way (ie define
standard metres and standard seconds and orthogonal reference frames
etc for purely local use) and from the primitive and \emph{definitive}
knowledge that, in the universe of our experience, the angular sizes
of distant objects are (statistically) in a strong inverse relationship
with measured cosmological redshifts (a redshift measurement is a
local measurement). A purely \emph{deductive} argument - containing
no additional hypothesese (beyond the simplifying, but strictly unnecessary,
assumption that the large scale universe is statistically unchanging
over all epochs) - then leads to the conclusion that, on very large
scales, space necessarily tends to become Euclidean and is supported
by a quasi-fractal $D\approx2$ distribution of universal material.

\subsection{Time in a Leibnitzian physical space: overview}

The notion of \emph{time} in the Leibnitzian universe is developed
in \S\ref{sec:The-temporal-dimension}, \S\ref{sec:Dynamical-constraints-in},
and \S\ref{sec:Physical-time}. The assumption that the universe
is statistically unchanging over all epochs is equivalent to stating
that there is no cosmic time. Therefore, to `do' dynamics in our Leibnitzian
universe some concept of local time must be introduced. To keep things
simple so that ideas can be properly clarified, we impose the requirement
that dynamics means Newtonian dynamics - in essence, that the Third
Law must be satisfied. This then leads to a natural definition of
time which, within the context of our Leibnitzian universe, can be
interpreted to mean that \emph{time} is an explicit measure change
in this universe\emph{,} which very much coincides with Mach's ideas
about the nature of time briefly mentioned in \S\ref{sub:A-brief-history}.

\subsection{Dynamics in the Leibnitzian universe: overview}

By dynamics here, we mean \emph{gravitational dynamics} and this is
discussed in \S\ref{sec:The-cosmological-potential} and \S\ref{sec:The-fractal,}.
We show that, on very large scales, all motions tend to become inertial
so that, when combined with the result that this large-scale Leibnitzian
universe is supported by a quasi-fractal $D\approx2$ material distribution,
the over-all picture is a close reflection of what we actually observe
on very large scales. \\
 \\
 On small scales in this universe, it becomes clear how Newtonian
gravitation emerges in a natural way - although we do not discuss
this in detail here. Of equal interest is the fact that for particular
limiting values of two parameters, the material in the universe becomes
exacty fractal, $D=2$, and and there is global dynamical equilibrium
on all scales. The significance of this is discussed.

\section{Leibnitzian physical space\label{sec:From-Leibnitz-to}}

\subsection{The qualitative model\label{sub:The-general-argument}}

In pursuit of the general question raised in the opening paragraph
of \S\ref{sec:Overview}, we shall assume an idealized universe:

\begin{itemize}
\item \emph{within which observers can define orthogonal frames locally
and make quantitative length} \emph{measurements/comparisons on local
scales only and make measurements of angular diameter and angular
position for objects at arbitrary distances;}  
\item \emph{which, on very large scales, consists of an infinite set of
identical discrete galaxies on which there is a primitive property
of absolute ordering which allows us to say only that galaxy $\mathcal{G}_{0}$
is} nearer/further \emph{than galaxy $\mathcal{G}_{1}$. There are
two obvious consequences of this latter requirement:}

\begin{itemize}
\item \emph{we must be able to see the galaxies and so they must emit light;} 
\item \emph{there must be some measurable parameter which allows the judgement
of} nearer/further\emph{. This could be, for example, measured angular
diameters or redshifts.}  
\end{itemize}
\item \emph{within which there is a primitive temporal ordering property
which allows a distinction to be made between} before \emph{and} after.
\emph{This property will provide a qualitative measure of} \textit{\emph{process}}
\emph{or} \textit{\emph{ordered change}} \emph{in the model universe;}
 
\item \emph{within which there is at least one origin about which the distribution
of `galaxies' is statistically isotropic - meaning that the results
of sampling along arbitrary lines of sight over sufficiently long
characteristic 'times' are independent of the directions of lines
of sight;}  
\item \emph{within which the distribution of `galaxies' is statistically
stationary - meaning that the results of sampling along arbitrary
lines of sight over sufficiently long characteristic 'times' are independent
of sampling epoch.} 
\end{itemize}

\subsubsection{Astrophysical spheres and a mass-calibrated metric for radial displacements}

We wish to obtain a conception of space \& time within which it is
\emph{meaningless} to talk about \emph{metric} in the absence of matter.
We imagine that we are in the world of our experience and that this
world is represented by the primitive model described above. Thus,
for the sake of definiteness, we will suppose that the primitive spatial
ordering of galaxies can be determined by redshift measurements -
in other words, we use only the knowledge that there exists a positive
correlation between the (statistical) angular diameters of galaxies
and their measured redshifts. \\
 \\
 It is then useful to discuss, briefly, the notion of spherical volumes
defined on large astrophysical scales in this universe: whilst we
can certainly give various precise operational definitions of spherical
volumes on local scales, the process of giving such definitions on
large scales is decidedly ambiguous. In effect, we have to suppose
that redshift measurements are (statistically) isotropic when taken
from an arbitrary point within the universe and that they vary monotonically
with distance on the large scales we are concerned with. With these
assumptions, spherical volumes can be defined (statistically) in terms
of redshift measurements - however, their radial calibration in terms
of ordinary units (such as metres) becomes increasingly uncertain
(and even unknown) on very large redshift scales.\\
 \\
 With these ideas in mind, the primary step taken in answer to the
elemental question of \S\ref{sub:The-general-argument} is the recognition
that, on large enough scales in the universe of our experience (say
$>30\, Mpc$), the amount of matter in a given redshift-defined spherical
volume in a given epoch can be considered as a well-defined (monotonic)
function of the sphere's (redshift) radius.

\emph{It follows immediately that a redshift calibration of the radius
of an astrophysical sphere has an equivalent mass-defined calibration.}

That is, the redshift-radius, $R_{z}$ say, of any spherical volume,
calibrated in terms of mass, can be considered given by: \begin{equation}
R_{z}=f(m)\rightarrow\delta R_{z}\approx f(m+\delta m)-f(m)\label{eqn0}\end{equation}
 where $m$ is the mass concerned and $f$ is some well-defined, but
unknown, monotonic increasing function of $m$. Thus, whatever the
form of $f$, we have immediately defined an invariant radial measurement
such that it becomes undefined in the absence of matter - in effect,
we have, in principle, a metric which follows Leibnitz in the required
sense for any displacement which is \emph{purely radial.}\\
 \\
 At this point, we can note that there will be some monotonic relation
between a redshift-defined radial displacement $R_{z}$ and a `conventional'
radial measure, $R$ say. Thus, the first of (\ref{eqn0}) above can
be written as\begin{equation}
R_{z}\equiv g(R)=f(m)\rightarrow R=g^{-1}(R_{z}),\,\, m=M(R)\leftrightarrow R=M^{-1}(m)\label{eqn0b}\end{equation}
 where $f$ is a well-defined, but unknown function of $m$, and $g$
is an arbitrary monotonic increasing function of $R$, the definition
of which provides a calibration of $R$ in terms of $R_{z}$.

\subsubsection{Definition of a three-dimensional coordinate frame}

We now consider how to generalize the foregoing ideas to provide coordinate
positions in three-dimensional space for arbitrarily placed distant
objects. The method is straighforward: since we have assumed that
an observer in the model universe can define a \emph{local} orthogonal
frame then he can set up a local rectangular frame with uncalibrated
axes, $\left(x^{1},x^{2},x^{3}\right)$ say, within which he can measure
the angular position of arbitrarily placed objects, $\left(\theta,\phi\right)$
say. These axes can be supposed extended indefinitely along the lines
of sight - but they are completely uncalibrated and cannot be assumed
orthogonal on large scales. However, since we have $R=g^{-1}(R_{z})$
where $R_{z}$ is a measured astrophysical quantity, then the observer
can provide a global calibration of his (locally) rectangular axes
to within a definition of $g$ by \emph{defining} \begin{equation}
x^{1}=R\sin\theta\cos\phi,\,\,\, x^{2}=R\sin\theta\sin\phi,\,\,\, x^{3}=R\cos\theta.\label{eqn0a}\end{equation}
 This procedure has the consequence that each of the three axes is
calibrated on an astrophysical scale in the same way as $R$ (that
is, to within a definition of $g$) and of guaranteeing that\[
R^{2}=\left(x^{1}\right)^{2}+\left(x^{2}\right)^{2}+\left(x^{3}\right)^{2}\rightarrow RdR=x^{i}dx^{j}\delta_{ij}.\]

\subsubsection{A mass-calibrated metric for arbitrary spatial displacements}

We are now in a position to generalize the idea of a mass-calibrated
radial displacement to that of a mass-calibration for arbitrary displacements
in our model universe. To this end, we use $m=M(R)$ from (\ref{eqn0b})
and use it in conjunction with (\ref{eqn0a}) to get the mass model:
\[
Mass\equiv m=M(R)\equiv M(x^{1},x^{2},x^{3}),\]
 for our rudimentary universe. \\
 \\
 Now consider the normal gradient vector $n_{a}=\nabla_{a}M$ and
the change in this arising from a displacement $dx^{k}$, \begin{equation}
dn_{a}=\nabla_{i}\left(\nabla_{a}M\right)\, dx^{i}\,,\label{eqn0A}\end{equation}
 where we assume that the geometrical connections required to give
this latter expression an unambiguous meaning are the usual metrical
connections - except of course, the metric tensor $g_{ab}$ of our
three-space is not yet defined.\\
 \\
 Given that $\nabla_{a}\nabla_{b}M$ is nonsingular, then (\ref{eqn0A})
provides a 1:1 mapping between the contravariant vector $dx^{a}$
and the covariant vector $dn_{a}$ so that, in the absence of any
other definition, we can \emph{define} $dn_{a}$ to be the covariant
form of $dx^{a}$. In this latter case the metric tensor of our three-space
automatically becomes\begin{equation}
g_{ab}=\nabla_{a}\nabla_{b}M\label{eqn0B}\end{equation}
 which, through the implied metrical connections, is a highly non-linear
equation defining $g_{ab}$ to within the specification of the mass
model, $M$. The scalar product $dS^{2}\equiv dn_{i}dx^{i}\equiv g_{ij}dx^{i}dx^{j}$
then provides an invariant measure for the magnitude of the infinitessimal
three-space displacement, $dx^{a}$.\\
 \\
 The units of $dS^{2}$ are easily seen to be those of $mass$ only
and so, in order to make them those of $length^{2}$ - as dimensional
consistency requires - we define the working invariant as $ds^{2}\equiv(2r_{0}^{2}/m_{0})dS^{2}$,
where $r_{0}$ and $m_{0}$ are scaling constants for the distance
and mass scales respectively and the numerical factor has been introduced
for later convenience. \\
 \\
 Finally, if we want \begin{equation}
ds^{2}\equiv\left(\frac{r_{0}^{2}}{2m_{0}}\right)dn_{i}dx^{i}\equiv\left(\frac{r_{0}^{2}}{2m_{0}}\right)g_{ij}dx^{i}dx^{j}\label{1a}\end{equation}
 to behave sensibly in the sense that $ds^{2}>0$ whenever $\left|d\textbf{r}\right|>0$
and $ds^{2}=0$ only when $\left|d\textbf{r}\right|=0$, then we must
replace the condition of non-singularity of $g_{ab}$ by the condition
that it is strictly positive definite; in the physical context of
the present problem, this will be considered to be a self-evident
requirement.

\subsection{The connection coefficients}

We have assumed that the geometrical connection coefficients can be
defined in some sensible way. To do this, we simply note that, in
order to define conservation laws (ie to do physics) in a Riemannian
space, it is necessary to be have a generalized form of Gausses' divergence
theorem in the space. This is certainly possible when the connections
are defined to be the metrical connections, but it is by no means
clear that it is ever possible otherwise. Consequently, the connections
are assumed to be metrical and so $g_{ab}$, given at (\ref{eqn0B}),
can be written explicitly as \begin{equation}
g_{ab}\equiv\nabla_{a}\nabla_{b}M\equiv\frac{\partial^{2}M}{\partial x^{a}\partial x^{b}}-\Gamma_{ab}^{k}\frac{\partial M}{\partial x^{k}},\label{(3)}\end{equation}
 where $\Gamma_{ab}^{k}$ are the Christoffel symbols, and given by
\[
\Gamma_{ab}^{k}~=~\frac{1}{2}g^{kj}\left(\frac{\partial g_{bj}}{\partial x^{a}}+\frac{\partial g_{ja}}{\partial x^{b}}-\frac{\partial g_{ab}}{\partial x^{j}}\right).\]

\section{The metric tensor given in terms of the mass model\label{sec:The-metric-tensor}}

\label{sec.6} It is shown, in appendix \ref{app.A}, how, for an
arbitrarily defined mass model, $m\equiv M(R)$, (\ref{(3)}) can
be exactly resolved to give an explicit form for $g_{ab}$ in terms
of such a general $M(R)$: Defining \[
\mathbf{R}\equiv(x^{1},x^{2},x^{3}),~~\Phi\equiv\frac{1}{2}\left(\mathbf{R\cdot R}\right)=\frac{1}{2}R^{2}~~{\rm and}~~M'\equiv\frac{dM}{d\Phi}\]
 then it is found that \begin{equation}
g_{ab}=A\delta_{ab}+Bx^{i}x^{j}\delta_{ia}\delta_{jb},\label{(4)}\end{equation}
 where \[
A\equiv\frac{d_{0}M+m_{1}}{\Phi},~~~B\equiv-\frac{A}{2\Phi}+\frac{d_{0}M'M'}{2A\Phi}.\]
 for arbitrary constants $d_{0}$ and $m_{1}$ where, as inspection
of the structure of these expressions for $A$ and $B$ shows, $d_{0}$
is a dimensionless scaling factor and $m_{1}$ has dimensions of mass.
Noting now that $M$ always occurs in the form $d_{0}M+m_{1}$, it
is convenient to write $\mathcal{M}\equiv d_{0}M+m_{1}$, and to write
$A$ and $B$ as \begin{equation}
A\equiv\frac{\mathcal{M}}{\Phi},~~B\equiv-\left(\frac{\mathcal{M}}{2\Phi^{2}}-\frac{\mathcal{M}'\mathcal{M}'}{2d_{0}\mathcal{M}}\right).\label{4a}\end{equation}

\section{An invariant calibration of the radial scale\label{sec:An-invariant-calibration}}

\label{sec.7} So far, we have assumed an arbitrary calibration for
the radial scale; that is, since, by (\ref{eqn0b}), $R_{z}=f(m)$
for unknown $f$ and since $R=g^{-1}(R_{z})$ for arbitrary monotonic
$g$, then $R=g^{-1}\left\{ f(m)\right\} $ is uncalibrated. In the
following, we show that the \emph{geodesic radial scale} provides
a unique calibration for $R$ in terms of $m$ and leads directly
to an understanding of fractality, $D\approx2$, in our observed universe.

\subsection{The geodesic radial scale}

Using (\ref{(4)}) and (\ref{4a}) in (\ref{1a}), and applying the
identities $x^{i}dx^{j}\delta_{ij}\equiv RdR$ and $\Phi\equiv R^{2}/2$,
we find, for an arbitrary displacement $d{\bf x}$, the invariant
measure: \[
ds^{2}=\left(\frac{R_{0}^{2}}{2m_{0}}\right)\left\{ \frac{\mathcal{M}}{\Phi}dx^{i}dx^{j}\delta_{ij}-\Phi\left(\frac{\mathcal{M}}{\Phi^{2}}-\frac{\mathcal{M}'\mathcal{M}'}{d_{0}\mathcal{M}}\right)dR^{2}\right\} ,\]
 which is valid for the arbitrary calibration $R=g^{-1}\left\{ f(m)\right\} $.
If the displacement $d{\bf x}$ is now constrained to be purely radial,
then we find \[
ds^{2}=\left(\frac{R_{0}^{2}}{2m_{0}}\right)\left\{ \Phi\left(\frac{\mathcal{M}'\mathcal{M}'}{d_{0}\mathcal{M}}\right)dR^{2}\right\} .\]
 Use of $\mathcal{M}'\equiv d\mathcal{M}/d\Phi$ and $\Phi\equiv R^{2}/2$
reduces this latter relationship to \begin{eqnarray*}
ds^{2} & = & \frac{R_{0}^{2}}{d_{0}m_{0}}\left(d\sqrt{\mathcal{M}}\right)^{2}~~\rightarrow~~ds=\frac{R_{0}}{\sqrt{d_{0}m_{0}}}d\sqrt{\mathcal{M}}~~\rightarrow\\
s & = & \frac{R_{0}}{\sqrt{d_{0}m_{0}}}\left(\sqrt{\mathcal{M}}-\sqrt{\mathcal{M}_{0}}\right),~~~{\rm where}~~~\mathcal{M}_{0}\equiv\mathcal{M}(s=0)\end{eqnarray*}
 which defines the invariant magnitude of an arbitrary \textit{radial}
displacement from the origin purely in terms of the mass-model representation
$\mathcal{M}\equiv d_{0}M+m_{1}$. By definition, this $s$ is the
invariant measure for an arbitrary finite radial displacement - in
other words, it is the natural measure of radial displacement. Thus,
by setting $R\equiv s$, we identify $R$ with the natural measure
of radial displacement, and thereby provide it with a unique calibration.\\
 \\
 To summarize, the natural mass-defined calibration for the radial
scale is given by \begin{equation}
R=\frac{R_{0}}{\sqrt{d_{0}m_{0}}}\left(\sqrt{\mathcal{M}}-\sqrt{\mathcal{M}_{0}}\right),\label{4e}\end{equation}
 where $\mathcal{M}_{0}$ is the value of $\mathcal{M}$ at $R=0$.

\subsection{The Euclidean metric and a fractal $D=2$ mass distribution }

\label{Euclidean} Using $\mathcal{M}\equiv d_{0}M+m_{1}$ and noting
that $M(R=0)=0$ necessarily, then $\mathcal{M}_{0}=m_{1}$ and so
(\ref{4e}) can be equivalently arranged as \begin{equation}
\mathcal{M}=\left[\frac{\sqrt{d_{0}m_{0}}}{R_{0}}R+\sqrt{m_{1}}\right]^{2}.\label{4b}\end{equation}
 Using $\mathcal{M}\equiv d_{0}M+m_{1}$ again, then the mass-distribution
function can be expressed in terms of the invariant radial displacement
as \begin{equation}
M=m_{0}\left(\frac{R}{R_{0}}\right)^{2}+2\sqrt{\frac{m_{0}m_{1}}{d_{0}}}\left(\frac{R}{R_{0}}\right)\label{4c}\end{equation}
 which, for the particular case $m_{1}=0$ becomes $M\,=\, m_{0}(R/R_{0})^{2}$.
Reference to (\ref{(4)}) shows that, with this mass distribution
and $d_{0}=1$, then $g_{ab}=\delta_{ab}$ so that the three-space
space becomes ordinary Euclidean space. Thus, whilst we have yet to
show that a globally inertial space can be associated with a non-trivial
global matter distribution (since no temporal dimension, and hence
no dynamics, has been introduced), we have shown that a globally \textit{Euclidean}
space can be associated with a non-trivial matter distribution, and
that this distribution is necessarily fractal with $D\,=\,2$.\\
 \\
 Note also that, on a large enough scale and for \textit{arbitrary}
values of $m_{1}$, (\ref{4c}) shows that radial distance varies
as the square-root of mass from the chosen origin - or, equivalently,
the mass varies as $R^{2}$. Consequently, on sufficiently large scales
Euclidean space is irreducibly related to a quasi-fractal, $D\,=\,2$,
matter distributions. Since $M/R^{2}\,\approx\, m_{0}/R_{0}^{2}$
on a large enough scale then, for the remainder of this paper, the
notation $g_{0}\equiv m_{0}/R_{0}^{2}$ is employed.

\section{The temporal dimension\label{sec:The-temporal-dimension}}

So far, the concept of `time' has only entered the discussion in a
qualitative was in \S\ref{sub:A-brief-history} - it has not entered
in any quantitative way and, until it does, there can be no discussion
of dynamical processes.\\
 \\
 Since, in its most general definition, time is a parameter which
orders change within a system, then a necessary pre-requisite for
its quantitative definition is a notion of change within the universe.
The most simple notion of change which can be defined in the universe
is that of changing relative spatial displacements of the objects
within it. Since our model universe is populated solely by identical
primitive `galaxies' then, in effect, all change is gravitational
change. This fact is incorporated into the cosmology to be derived
by constraining all particle displacements to satisfy the Weak Equivalence
Principle. We are then led to a Lagrangian description of particle
motions in which the Lagrange density is degree zero in its temporal-ordering
parameter. From this, it follows that the corresponding Euler-Lagrange
equations form an \textit{incomplete} set. \\
 \\
 The origin of this problem traces back to the fact that, because
the Lagrangian density is degree zero in the temporal ordering parameter,
it is then invariant with respect to any transformation of this parameter
which preserves the ordering. This implies that, in general, temporal
ordering parameters cannot be identified directly with physical time
- they merely share one essential characteristic. This situation is
identical to that encountered in the Lagrangian formulation of General
Relativity; there, the situation is resolved by defining the concept
of {\lq particle proper time'}. In the present case, this is not
an option because the notion of particle proper time involves the
prior definition of a system of observer's clocks - so that some notion
of clock-time is factored into the prior assumptions upon which General
Relativity is built.\\
 \\
 In the present case, it turns out that the isotropies already imposed
on the system conspire to provide an automatic resolution of the problem
which is consistent with the already assumed interpretation of {\lq
time'} as a measure of ordered change in the model universe. To be
specific, it turns out that the elapsed time associated with any given
particle displacement is proportional, via a scalar field, to the
invariant spatial measure attached to that displacement. Thus, physical
time is defined directly in terms of the invariant measures of \textit{process}
within the model universe and, furthermore, local conditions affect
clock rates.

\section{Equations of motion for generalized time parameter\label{sec:Dynamical-constraints-in}}

\label{sec.constr} Firstly, and as already noted, we are assuming
that, within our model universe, all motions are gravitational, and
we model this circumstance by constraining all such motions to satisfy
the Weak Equivalence Principle by which we mean that the trajectory
of a body is independent of its internal constitution. This constraint
can be expressed as:

\textbf{\emph{C1}} \emph{Particle trajectories are independent of
the specific mass values of the particles concerned;}  

\noindent Secondly, given the isotropy conditions imposed on the model
universe from the chosen origin, symmetry arguments lead to the conclusion
that the net action of the whole universe of particles acting on any
given single particle is such that any net acceleration of the particle
must always appear to be directed through the coordinate origin. Note
that this conclusion is \textit{independent} of any notions of retarded
or instantaneous action. This constraint can then be stated as:

\textbf{\emph{C2}} \emph{Any acceleration of any given material particle
must necessarily be along the line connecting the particular particle
to the coordinate origin.}  

Now suppose $p$ and $q$ are two arbitrarily chosen point coordinates
on the trajectory of the chosen particle, and suppose that (\ref{1a})
is integrated between these points to give the scalar invariant \begin{equation}
I(p,q)=\int_{p}^{q}\left(\frac{1}{\sqrt{2g_{0}}}\right)\sqrt{dn_{i}dx^{i}}\equiv\int_{p}^{q}\left(\frac{1}{\sqrt{2g_{0}}}\right)\sqrt{g_{ij}dx^{i}dx^{j}}.\label{(2)}\end{equation}
 Then, in accordance with the foregoing interpretation, $I(p,q)$
gives a scalar record of how the particle has moved between $p$ and
$q$ defined with respect to the particle's continually changing relationship
with the mass model, $M(R)$.\\
 \\
 If $I(p,q)$ is now minimized with respect to choice of the trajectory
connecting $p$ and $q$, then this minimizing trajectory can be interpreted
as a geodesic in the Riemannian space which has $g_{ab}$ as its metric
tensor. Given that $g_{ab}$ is defined in terms of the mass model
$M(R)$ - the existence of which is independent of any notion of `inertial
mass', then the existence of the metric space, and of geodesic curves
within it, is likewise explicitly independent of any concept of inertial-mass.
It follows that the identification of the particle trajectory ${\bf r}$
with these geodesics means that particle trajectories are similarly
independent of any concept of inertial mass, and can be considered
as the modelling step defining that general subclass of trajectories
which conform to that characteristic phenomenology of gravitation
defined by condition \textbf{C1} above.\\
\\
The geodesic equations in the space with the metric tensor (\ref{(4)})
can now be obtained, in the usual way, by defining the Lagrangian
density \begin{equation}
{\cal L}\equiv\left(\frac{1}{\sqrt{2g_{0}}}\right)\sqrt{g_{ij}\dot{x}^{i}\dot{x}^{j}}=\left(\frac{1}{\sqrt{2g_{0}}}\right)\left(A\left(\mathbf{\dot{R}\cdot\dot{R}}\right)+B\dot{\Phi}^{2}\right)^{1/2},\label{4d}\end{equation}
 where $\dot{x}^{i}\equiv dx^{i}/dt$, etc., and writing down the
Euler-Lagrange equations \begin{eqnarray}
2A\mathbf{\ddot{R}} & + & \left(2A'\dot{\Phi}-2\frac{\dot{{\cal L}}}{{\cal L}}A\right)\mathbf{\dot{R}}+\left(B'\dot{\Phi}^{2}+2B\ddot{\Phi}-A'\left(\mathbf{\dot{R}\cdot\dot{R}}\right)-2\frac{\dot{{\cal L}}}{{\cal L}}B\dot{\Phi}\right)\mathbf{R}\nonumber \\
 & = & 0,\label{(5)}\end{eqnarray}
 where $\mathbf{\dot{R}}\equiv d\mathbf{R}/dt$ and $A'\equiv dA/d\Phi$,
etc. By identifying particle trajectories with geodesic curves, this
equation is now interpreted as the equation of motion, referred to
the chosen origin, of a single particle satisfying condition \textbf{C1}
above.\\
 \\
 However, noting that the variational principle, (\ref{(2)}), is
of order zero in its temporal ordering parameter, we can conclude
that the principle is invariant with respect to arbitrary transformations
of this parameter; in turn, this means that the temporal ordering
parameter cannot be identified with physical time. This problem manifests
itself formally in the statement that the equations of motion (\ref{(5)})
do not form a complete set, so that it becomes necessary to specify
some extra condition to close the system.\\
 \\
 A similar circumstance arises in General Relativity when the equations
of motion are derived from an action integral which is formally identical
to (\ref{(2)}). In that case, the system is closed by specifying
the arbitrary time parameter to be the {\lq proper time'}, so that
\begin{equation}
d\tau={\cal L}(x^{j},dx^{j})~~\rightarrow~~{\cal L}(x^{j},\frac{dx^{j}}{d\tau})=1,\label{(5a)}\end{equation}
 which is then considered as the necessary extra condition required
to close the system. In the present circumstance, we are rescued by
the, as yet, unused condition \textbf{C2}.

\section{Physical time\label{sec:Physical-time}}

\subsection{Completion of equations of motion}

Consider \textbf{C2}, which states that any particle accelerations
must necessarily be directed through the coordinate origin. This latter
condition simply means that the equations of motion must have the
general structure \[
\mathbf{\ddot{R}}=G(t,\mathbf{R},\mathbf{\dot{R}})\mathbf{R},\]
 for scalar function $G(t,\mathbf{R},\mathbf{\dot{R}})$. In other
words, (\ref{(5)}) satisfies condition \textbf{C2} if the coefficient
of $\mathbf{\dot{R}}$ is zero, so that \begin{equation}
\left(2A'\dot{\Phi}-2\frac{\dot{{\cal L}}}{{\cal L}}A\right)=0~~~\rightarrow\frac{A'}{A}\dot{\Phi}=\frac{\dot{{\cal L}}}{{\cal L}}~~\rightarrow~~{\cal L}=k_{0}A,\label{(6)}\end{equation}
 for arbitrary constant $k_{0}$ which is necessarily positive since
$A>0$ and ${\cal L}>0$. The condition (\ref{(6)}), which guarantees
(\textbf{C2}), can be considered as the condition required to close
the incomplete set (\ref{(5)}) and is directly analogous to (\ref{(5a)}),
the condition which defines `proper time' in General Relativity.

\subsection{Physical time defined as process}

\label{sec.9a} Equation (\ref{(6)}) can be considered as that equation
which removes the pre-existing arbitrariness in the {\lq time'}
parameter by \textit{defining} physical time:- from (\ref{(6)}) and
(\ref{4d}) we have \begin{eqnarray}
{\cal L}^{2} & = & k_{0}^{2}A^{2}~\rightarrow~A\left(\mathbf{\dot{R}\cdot\dot{R}}\right)+B\dot{\Phi}^{2}=2g_{0}k_{0}^{2}A^{2}~\rightarrow~\nonumber \\
g_{ij}\dot{x}^{i}\dot{x}^{j} & = & 2g_{0}k_{0}^{2}A^{2}\label{(7)}\end{eqnarray}
 so that, in explicit terms, physical time is \textit{defined} by
the relation \begin{equation}
dt^{2}=\left(\frac{1}{2g_{0}k_{0}^{2}A^{2}}\right)g_{ij}dx^{i}dx^{j},~~~{\rm where}~~A\equiv{\frac{{\cal M}}{\Phi}}.\label{(8)}\end{equation}
 In short, the elapsing of time is given a direct physical interpretation
in terms of the process of \textit{displacement} in the model universe.\\
 \\
 Finally, noting that, by (\ref{(8)}), the dimensions of $k_{0}^{2}$
are those of $L^{6}/[T^{2}\times M^{2}]$, then the fact that $g_{0}\equiv m_{0}/R_{0}^{2}$
(cf \S\ref{sec.7}) suggests the change of notation $k_{0}^{2}\propto v_{0}^{2}/g_{0}^{2}$,
where $v_{0}$ is a constant having the dimensions (but not the interpretation)
of {\lq velocity'}. So, as a means of making the dimensions which
appear in the development more transparent, it is found convenient
to use the particular replacement $k_{0}^{2}\equiv v_{0}^{2}/(4d_{0}^{2}g_{0}^{2})$.
With this replacement, the \textit{definition} of physical time, given
at (\ref{(8)}), becomes \begin{equation}
dt^{2}=\left(\frac{4d_{0}^{2}g_{0}}{v_{0}^{2}A^{2}}\right)g_{ij}dx^{i}dx^{j}.\label{(8a)}\end{equation}
 Since, as is easily seen from the definition of $g_{ab}$ given in
\S\ref{sec.6}, $g_{ij}dx^{i}dx^{j}$ is necessarily finite and non-zero
for a non-trivial displacement $d\mathbf{R}$

\subsection{The necessity of $v_{0}^{2}\,\neq\,0$}

\label{sec.12.3} Equation (\ref{(8a)}) provides a definition of
physical time in terms of basic process (displacement) in the model
universe. Since the parameter $v_{0}^{2}$ occurs nowhere else, except
in its explicit position in (\ref{(8a)}), then it is clear that setting
$v_{0}^{2}=0$ is equivalent to physical time becoming undefined.
Therefore, of necessity, $v_{0}^{2}\neq0$ and all non-zero finite
displacements are associated with a non-zero finite elapsed physical
time.

\section{The cosmological potential\label{sec:The-cosmological-potential}}

The model is most conveniently interpreted when expressed in potential
terms and so, in the following, it is shown how this is done.

\subsection{The equations of motion: potential form}

\label{sec.10} When (\ref{(6)}) is used in (\ref{(5)}) we get:\begin{equation}
2A\mathbf{\ddot{R}}+\left(B'\dot{\Phi}^{2}+2B\ddot{\Phi}-A'\left(\mathbf{\dot{R}\cdot\dot{R}}\right)-2\frac{A'}{A}B\dot{\Phi}^{2}\right)\mathbf{R}=0.\label{(9)}\end{equation}
 Suppose we define a function $V$ according to $V\equiv C_{0}-\left(\mathbf{\dot{R}\cdot\dot{R}}\right)/2$,
for some arbitrary constant $C_{0}$; then, by (\ref{(7)}) \begin{equation}
V\equiv C_{0}-\frac{1}{2}\left(\mathbf{\dot{R}\cdot\dot{R}}\right)=C_{0}-\frac{v_{0}^{2}}{4d_{0}^{2}g_{0}}A+\frac{B}{2A}\dot{\Phi}^{2},\label{(10)}\end{equation}
 where $A$ and $B$ are defined at (\ref{4a}). With unit vector,
$\mathbf{\hat{R}}$, then appendix {\ref{app.C} shows how this function
can be used to express (\ref{(9)}) in the potential form \begin{equation}
\mathbf{\ddot{R}}=-\frac{dV}{dr}\mathbf{\hat{R}}\label{(11)}\end{equation}
 so that $V$ is a potential function, and $C_{0}$ is the arbitrary
constant usually associated with a potential function.

\subsection{The potential function, $V$, as a function of $R$}

\label{sec.centres} From (\ref{(10)}), we have \[
2C_{0}\,-\,2V\,=\,\dot{R}^{2}+R^{2}\dot{\theta}^{2}=\frac{v_{0}^{2}}{2d_{0}^{2}g_{0}}A-\frac{B}{A}R^{2}\dot{R}^{2}\]
 so that $V$ is effectively given in terms of $R$ and $\dot{R}$.
In order to clarify things further, we now eliminate the explicit
appearance of $\dot{R}$. Since all forces are central, then angular
momentum is conserved; consequently, after using conserved angular
momentum, $h$, and the definitions of $A$, $B$ and ${\cal M}$
given in \S\ref{sec.6}, the foregoing equations can be written as
\begin{eqnarray}
2C_{0} & - & 2V~~\,=\,\nonumber \\
\dot{R}^{2} & + & R^{2}\dot{\theta}^{2}\,=\, v_{0}^{2}+\frac{4v_{0}^{2}}{R}\sqrt{\frac{m_{1}}{d_{0}g_{0}}}+\frac{d_{0}-1}{R^{2}}\left(\frac{6m_{1}v_{0}^{2}}{d_{0}^{2}\, g_{0}}-h^{2}\right)\nonumber \\
 & + & \frac{2}{R^{3}}\sqrt{\frac{d_{0}m_{1}}{g_{0}}}\left(\frac{2m_{1}v_{0}^{2}}{d_{0}^{2}\, g_{0}}-h^{2}\right)+\frac{1}{R^{4}}\frac{m_{1}}{g_{0}}\left(\frac{m_{1}v_{0}^{2}}{d_{0}^{2}\, g_{0}}-h^{2}\right)\label{11e}\end{eqnarray}
 so that $V(R)$ is effectively given by the right-hand side of (\ref{11e}).
Various interesting features of this potential function, which go
beyond our immediate interest, are discussed in appendix \ref{app.B}.

\section{The fractal $D\,$=$\,2$ inertial universe\label{sec:The-fractal,}}

The model we have developed allows either a perfectly fractal $D=2$
inertial universe as a specific limiting case, or a quasi-fractal
$D\approx2$, almost inertial universe on sufficiently large scales.
In either case, the notion of an empty metrical universe is meaningless,
so that the question originally posed in \S\ref{sec:Overview} is
finally answered. We consider each case in turn.

\subsection{The perfectly fractal $D=2$ universe}

\label{Fractal} Reference to (\ref{11e}) shows that the parameter
choice $m_{1}=0$ and $d_{0}=1$ makes the potential function constant
everywhere so that there is a global dynamical equilibrium; similarly,
(\ref{4c}) shows how this equilibrium universe is supported by a
mass distribution which is necessarily distributed as an exact fractal
with $D=2$.\\
 \\
 A more detailed analysis is illuminating: specifically, for the case
$m_{1}=0$, $d_{0}=1$, the definition $M$ at (\ref{4c}) together
with the definitions of $A$ and $B$ in \S\ref{sec.6} give \[
A\,=\,\frac{2m_{0}}{R_{0}^{2}},~~~B\,=\,0\]
 so that, by (\ref{(10)}) (remembering that $g_{0}\equiv m_{0}/R_{0}^{2}$)
we have \begin{equation}
\left(\mathbf{\dot{R}\cdot\dot{R}}\right)\,=\, v_{0}^{2}\label{(12)}\end{equation}
 for all displacements in the model universe. It is (almost) natural
to assume that the constant $v_{0}^{2}$ in (\ref{(12)}) simply refers
to the constant velocity of any given particle, and likewise to assume
that this can differ between particles. However, each of these assumptions
would be wrong since - as we now show - $v_{0}^{2}$ is, firstly,
more properly interpreted as a conversion factor from spatial to temporal
units and, secondly, is a \textit{global} constant which applies equally
to all particles.\\
 \\
 To understand these points, we begin by noting that (\ref{(12)})
is a special case of (\ref{(7)}) and so, by (\ref{(8)}), is more
accurately written as \begin{equation}
dt^{2}\,=\,\frac{1}{v_{0}^{2}}\left(d\mathbf{R\cdot}d\mathbf{R}\right)\label{(13)}\end{equation}
 which, by the considerations of \S\ref{sec.9a}, we recognize as
the \textit{definition} of the elasped time experienced by any particle
undergoing a spatial displacement $d{\bf r}$ in the model inertial
universe. Since this universe is isotropic about all points, then
there is nothing which can distinguish between two separated particles
(other than their separateness) undergoing displacements of equal
magnitudes; consequently, each must be considered to have experienced
equal elapsed times. It follows from this that $v_{0}^{2}$ is not
to be considered as a locally defined particle velocity, but is a
\textit{globally} defined constant which has the effect of converting
between spatial and temporal units of measurement.\\
 \\
 We now see that the model inertial universe, with (\ref{(13)}) as
a global relationship, bears a close formal resemblance to a universe
filled purely with Einsteinien photons - the difference being, of
course, that the particles in the model inertial universe are assumed
to be countable and to have mass properties. This formal resemblance
means that the idealized inertial universe can be likened to a quasi-photon
fractal, $D=2$, gas universe.

\subsection{The quasi-fractal $D\approx2$ universe}

On large scales, and for arbitrary choice of the parameters, (\ref{4c})
shows that $M\sim R^{2}$ on sufficiently large scales whilst (\ref{11e})
shows that accelerations $\rightarrow0$ on these scales. Using arguments
similar to those used above, we can also show that the elapsed time
experienced by particles undergoing displacements in this universe
becomes a function of the displacement itself and the local environment.

\subsection{Implications for theories of gravitation}

Given that gravitational phenomena are usually considered to arise
as mass-driven perturbations of flat inertial backgrounds, then the
foregoing results - to the effect that the inertial background is
necessarily associated with a non-trivial fractal matter distribution
- must necessarily give rise to completely new perspectives about
the nature and properties of gravitational phenomena. However, such
considerations take us beyond the immediate concerns of this paper.

\section{Summary and Conclusions}

In practice, and insofar as we can determine, the universe of our
experience is asymptotically inertial on the very large scales at
which quasi-fractal $D\approx2$ behaviour appears to dominate. We
adopted the tentative hypothesis that this quasi-fractal behaviour
will be found to extend to all observable (large) scales and that
it is a persistant signature reflecting the fundamental nature of
the universe as a place in which the ideas Aristotle, Leibnitz, Berkeley
and Mach are a fair reflection of the reality. If this is the case,
then it should be possible to produce a theoretical description of
such a universe so that we were led to ask the question:

\textit{Is it possible to associate a globally inertial space \& time
with a non-trivial global matter distribution and, if it is, what
are the fundamental properties of this distribution?} 

We analysed the question in the context of a simple model universe
which was assumed to possess only the primitive concepts of \emph{nearer/further}
and of \emph{before/after.} We then showed how quantitative ideas
of spatial and temporal metrics could be defined entirely in terms
of the universal material and that, as a direct consequence:

\begin{itemize}
\item on sufficiently large scales in this universe and for an arbitrary
choice of the parameters, the distribution of material becomes quasi-fractal
with $D\approx2$ (cf equation (\ref{4b}) whilst accelerations tend
to zero (inertial behaviour on large scales, cf (\ref{11e})); 
\item In an ideal limiting case of two particular parameters taking certain
exact values, a globally inertial space \& time is irreducibly related
to a fractal, $D=2$, distribution of material within the model universe. 
\end{itemize}
However, the latter ideal inertial universe is distinguished in the
sense that whilst all the particles within it have arbitrarily directed
motions, the particle velocities all have \textit{equal} magnitude.
In this sense, this limiting case globally inertial model universe
is more accurately to be considered as a quasi-photon gas universe
than the universe of our macroscopic experience. In other words, it
looks more like a crude model of a material vacuum than the universe
of our direct experience.

\appendix
%dummy comment inserted by tex2lyx to ensure that this paragraph is not empty
%dummy comment inserted by tex2lyx to ensure that this paragraph is not empty
{}

\section{A Resolution of the Metric Tensor}

\label{app.A} The general system is given by \[
g_{ab}=\frac{\partial^{2}M}{\partial x^{a}\partial x^{b}}-\Gamma_{ab}^{k}\frac{\partial M}{\partial x^{k}},\]
 \[
\Gamma_{ab}^{k}~\equiv~\frac{1}{2}g^{kj}\left(\frac{\partial g_{bj}}{\partial x^{a}}+\frac{\partial g_{ja}}{\partial x^{b}}-\frac{\partial g_{ab}}{\partial x^{j}}\right),\]
 and the first major problem is to express $g_{ab}$ in terms of the
reference scalar, $M$. The key to this is to note the relationship
\[
\frac{\partial^{2}M}{\partial x^{a}\partial x^{b}}=M'\delta_{ab}+M''x^{a}x^{b},\]
 where $M'\equiv dM/d\Phi$, $M''\equiv d^{2}M/d\Phi^{2}$ and $\Phi\equiv R^{2}/2$,
since this immediately suggests the general structure \begin{equation}
g_{ab}=A\delta_{ab}+Bx^{a}x^{b},\label{A1}\end{equation}
 for unknown functions, $A$ and $B$. It is easily found that \[
g^{ab}=\frac{1}{A}\left[\delta_{ab}-\left(\frac{B}{A+2B\Phi}\right)x^{a}x^{b}\right]\]
 so that, with some effort, \[
\Gamma_{ab}^{k}=\frac{1}{2A}H_{1}-\left(\frac{B}{2A(A+2B\Phi)}\right)H_{2}\]
 where \begin{eqnarray*}
H_{1} & = & A'(x^{a}\delta_{bk}+x^{b}\delta_{ak}-x^{k}\delta_{ab})\\
 & + & B'x^{a}x^{b}x^{k}+2B\delta_{ab}x^{k}\end{eqnarray*}
 and \begin{eqnarray*}
H_{2} & = & A'(2x^{a}x^{b}x^{k}-2\Phi x^{k}\delta_{ab})\\
 & + & 2\Phi B'x^{a}x^{b}x^{k}+4\Phi Bx^{k}\delta_{ab}.\end{eqnarray*}
 Consequently, \begin{eqnarray*}
g_{ab} & = & {\frac{\partial^{2}M}{\partial x^{a}\partial x^{b}}}-\Gamma_{ab}^{k}{\frac{\partial M}{\partial x^{k}}}\equiv\delta_{ab}M'\left({\frac{A+A'\Phi}{A+2B\Phi}}\right)\\
 & + & x^{a}x^{b}\left(M''-M'\left({\frac{A'+B'\Phi}{A+2B\Phi}}\right)\right).\end{eqnarray*}
 Comparison with (\ref{A1}) now leads directly to \begin{eqnarray*}
A & = & M'\left({\frac{A+A'\Phi}{A+2B\Phi}}\right)=M'\left(\frac{(A\Phi)'}{A+2B\Phi}\right),\\
B & = & M''-M'\left({\frac{A'+B'\Phi}{A+2B\Phi}}\right).\end{eqnarray*}
 The first of these can be rearranged as \[
B={\frac{M'}{2\Phi}}\left(\frac{(A\Phi)'}{A}\right)-{\frac{A}{2\Phi}}\]
 or as \[
\left(\frac{M'}{A+2B\Phi}\right)={\frac{A}{(A\Phi)'}},\]
 and these expressions can be used to eliminate $B$ in the second
equation. After some minor rearrangement, the resulting equation is
easily integrated to give, finally, \[
A\equiv{\frac{d_{0}M+m_{1}}{\Phi}},~~~B\equiv-{\frac{A}{2\Phi}}+{\frac{d_{0}M'M'}{2A\Phi}}.\]

\section{Conservative Form of Equations of Motion}

\label{app.C} From (\ref{(10)}), we have \begin{equation}
V\equiv-\frac{1}{2}\left(\mathbf{\dot{R}\cdot\dot{R}}\right)=-\frac{k_{0}^{2}A}{2}+\frac{B}{2A}\dot{\Phi}^{2},\label{B1}\end{equation}
 from which we easily find \[
\frac{dV}{dR}\equiv\frac{\partial V}{\partial R}+\frac{\partial V}{\partial\dot{R}}\frac{\ddot{R}}{\dot{R}}\]
 \[
=\frac{-k_{0}^{2}A'}{2}R+\frac{\dot{\Phi}^{2}R}{2A}\left(B'-\frac{A'B}{A}\right)+\frac{B}{A}\left(R\dot{R}^{2}+R^{2}\ddot{R}\right)\]
 where we remember that $A'\equiv dA/d\Phi$ etc. Since $\dot{R}^{2}+R\ddot{R}=\ddot{\Phi}$,
then the above leads to \[
\frac{dV}{dR}\mathbf{\hat{R}}=\left(\frac{-k_{0}^{2}A'}{2}+\frac{B'}{2A}\dot{\Phi}^{2}-\frac{A'B}{2A^{2}}\dot{\Phi}^{2}+\frac{B}{A}\ddot{\Phi}\right)\mathbf{R}.\]
 Writing (\ref{(11)}) as \[
2A\mathbf{\ddot{R}}+2A\frac{dV}{dR}\mathbf{\hat{R}}=0,\]
 and using the above expression, we get the equation of motion as
\begin{equation}
2A\mathbf{\ddot{R}}+\left(-k_{0}^{2}AA'+B'{\dot{\Phi}}^{2}-{\frac{A'B}{A}}{\dot{\Phi}}^{2}+2B{\ddot{\Phi}}\right)\mathbf{R}=0.\label{B2}\end{equation}
 Finally, from (\ref{B1}), we have \[
k_{0}^{2}A={\frac{B}{A}}{\dot{\Phi}}^{2}+\left(\mathbf{\dot{R}\cdot\dot{R}}\right),\]
 which, when substituted into (\ref{B2}), gives (\ref{(9)}).

\section{Outline analysis of the potential function}

\label{app.B} It is quite plain from (\ref{11e}) that, for any $m_{1}\neq0$,
then the model universe has a preferred centre and that the parameter
$m_{1}$ (which has dimensions of mass) plays a role in the potential
$V$ which is analogous to the source mass in a Newtonian spherical
potential - that is, the parameter $m_{1}$ can be identified as the
mass of the potential source in the model universe. However, setting
$m_{1}=0$ is not sufficient to guarantee a constant potential field
since any $d_{0}\neq1$ also provides the model universe with a preferred
centre. The role of $d_{0}$ is most simply discussed in the limiting
case of $m_{1}=0$: in this case, the second equation of (\ref{11e})
becomes \begin{equation}
\dot{R}^{2}+R^{2}\dot{\theta}^{2}\,=\, v_{0}^{2}-(d_{0}-1)\frac{h^{2}}{R^{2}}.\label{11f}\end{equation}
 If $d_{0}<1$ then $|\mathbf{\dot{R}}|\rightarrow\infty$ as $R\rightarrow0$
so that a singularity exists. Conversely, remembering that $v_{0}^{2}>0$
(cf \S\ref{sec.12.3}) then, if $d_{0}>1$, equation (\ref{11f})
restricts real events to the exterior of the sphere defined by $R^{2}=(d_{0}-1)h^{2}/v_{0}^{2}$.
In this case, the singularity is avoided and the central {\lq massless
particle'} is given the physical property of {\lq finite extension'}.
In the more realistic case for which $m_{1}>0$, reference to (\ref{11e})
shows that the $R=0$ singularity is completely avoided whenever $h^{2}>m_{1}v_{0}^{2}/d_{0}^{2}g_{0}$
since then a {\lq finite extension'} property for the central massive
particle always exists. Conversely, a singularity will necessarily
exist whenever $h^{2}\leq m_{1}v_{0}^{2}/d_{0}^{2}g_{0}$.

In other words, the model universe has a preferred centre when either
$m_{1}>0$, in which case the source of the potential is a massive
central particle having various properties depending on the value
of $d_{0}$, or when $m_{1}=0$ and $d_{1}\neq0$.

\end{document}